# Engineered spatial inversion symmetry breaking in an oxide heterostructure built from isosymmetric room-temperature magnetically ordered components

J. Alaria[a,b], P. Borisov[a,f], M. S. Dyer[a], T. D. Manning[a], S. Lepadatu[c], M. G. Cain[c], E. D. Mishina[d], N. E. Sherstyuk[d], N.A. Ilyin[d], J. Hadermann[e], D. Lederman[f], J. B. Claridge[a]*, and M. J. Rosseinsky[a]*

The oxide heterostructure [(YFeO$_3$)$_5$(LaFeO$_3$)$_5$]$_{40}$, which is magnetically ordered and piezoelectric at room temperature, has been constructed from two weak ferromagnetic AFeO$_3$ perovskites with different A cations using RHEED-monitored pulsed laser deposition. The polarisation arises through the removal of inversion centres present within the individual AFeO$_3$ components. This symmetry reductionis a result of combining ordering on the A site, imposed by the periodicity of the grown structure, with appropriate orientations of the octahedral tilting characteristic of the perovskite units themselves, according to simple symmetry-controlled rules. The polarisation is robust against A site interdiffusion between the two layers which produces a sinusoidally modulated occupancy that retains the coupling of translational and point symmetries required to produce a polar structure. Magnetization and magneto-optical Kerr rotation measurements show that the heterostructure's magnetic structure is similar to that of the individual components. Evidence of the polarity was obtained from second harmonic generation and piezoelectric force microscopy measurements. Modeling of the piezoresponse allows extraction of d$_{33}$ (approximately 10 pC/N) of the heterostructure, which is in agreement with DFT calculations.

## Introduction

The breaking of inversion symmetry to generate a polarization is a prerequisite for the technologically important properties of ferro- and piezoelectricity used in capacitors and actuators, while the breaking of time-reversal symmetry is required for the magnetically ordered states such as antiferro- and ferromagnetism used in information storage. It is however chemically challenging to combine both properties in a single material, because there is a competition in the electronic structure requirements between many of the mechanisms responsible for forming each state[1] *e.g.* the crystal chemistry of the closed-shell $d^0$ $Ti^{4+}$ and $s^2$ $Pb^{2+}$cations driving ferroelectricity in $PbZr_{1-x}Ti_xO_3$ does not afford the unpaired electrons required for magnetic order. It is possible to approach this problem by combining the two chemistries required, as in $BiFeO_3$ where $Fe^{3+}$ provides antiferromagnetism and $Bi^{3+}$ ferroelectricity, by making composites over a range of length scales between compounds which individually display one of the two required ground states,[2] or by coupling reduced spatial symmetry with the onset of magnetic order.[3-5]

Oxide heterostructures display emergent phenomena controlled by charge, orbital and spin reconstruction at the internal interfaces within the thin films.[6] Inversion and time-reversal symmetries have been broken together in tricolor heterostructures employing multiple component compositions and symmetries[7], or by using substrate-induced strain.[8]

The $ABO_3$ perovskite structure supports both magnetic and polar ground states. The diverse array of physical properties can be controlled via tilting of the $BO_6$ octahedra[9] through the B-O-B overlap, and by ordering of cations on both the A[10, 11] and B sites[12]. Thin film heterostructures of $ABO_3$ materials afford new properties arising from strain and internal interfaces[6]. Recent theoretical work[13-16] has proposed that specific combinations of cation order and tilting, originally elucidated for HRTEM defect analysis of bulk materials[17], can impose polarity on $(ABO_3:A'BO_3)$ 1:1 heterostructures where both components adopt the Pnma structure. This involves out-of-phase octahedral tilting along two pseudocubic $a_p$ perovskite subcell directions denoted $a^-$ and in-phase tilting along the third, denoted $b^+$, which becomes the b axis of the $a_{Pnma} = \sqrt{2}a_p$, $b_{Pnma} = 2a_p$ and $c_{Pnma} = \sqrt{2}a_p$ unit cell (Figure 1). This mechanism is distinct from compositionally-generated polarity in tricolor superlattices[7, 18] as it operates in two-component isosymmetric systems through the coupling of point symmetry to translational compositional modulation. By increasing the separation between the polar interfaces while retaining the inversion symmetry breaking by coupling A site modulation to correctly oriented octahedral rotation, a polarity is generated in a heterostructure built from two magnetically ordered perovskites, thereby combining both broken space inversion and time-reversal symmetries at room temperature.

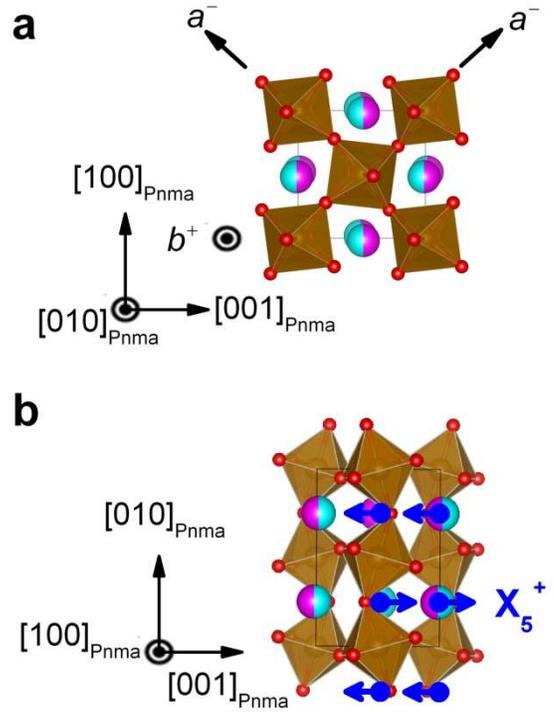

**Figure 1:** Projection of the Pnma $GdFeO_3$ structure (a) projected along the **b** axis showing the orientation of the tilt axes and (b), projected along **a** highlighting the $X_5^+$ A site antiferrodistortive displacement mode symmetry. Atoms are colored La (cyan), Y (magenta), Fe (brown) and O (red) throughout the manuscript.

## Experimental & computational details

$(YFeO_3)_5/(LaFeO_3)_5$, $(YFeO_3)_4/(LaFeO_3)_4$, $YFeO_3$, and $LaFeO_3$ films with a total thickness of 160 nm on atomically flat (101) $DyScO_3$ single crystal substrates and a 160 nm $(YFeO_3)_5/(LaFeO_3)_5$ film on atomically flat (001) $SrTiO_3$ were grown using pulsed laser deposition with a KrF(248nm) ultraviolet excimer laser. The $SrTiO_3$ and $DyScO_3$ substrates were prepared following the procedure described in the literature[19, 20]; the surface morphology of the treated substrate showed clear step and terraces structure (Fig. S3).The targets were prepared by solid state reaction of the dried oxides, the deposition was carried out under 0.7 mTorr of high purity $O_2$ with a substrate temperature of 700°C and a laser fluence of 1 J/cm$^2$ and a frequency of 3-5 Hz. The growth was monitored using a differentially pumped RHEED system by recording the specular reflection intensity variation.

X-ray diffraction was carried out using a Panalytical X'Pert Pro MRD equipped with a four bounce Ge (220) monochromator.

High angle annular dark field scanning transmission electron microscopy (HAADF-STEM) images were acquired on an FEI Titan 50–80 microscope equipped with a probe aberration corrector and operated at 300 kV. The sample was prepared with focused ion beam (FIB) on a Helios NanoLab 650.

The magnetization of the films grown on SrTiO$_3$ substrates was measured in a MPMS SQUID magnetometer (Quantum Design) at room temperature with the applied field parallel and perpendicular to the film surface.

Magnetic hysteresis loops of the films grown on DyScO$_3$ were recorded using the magneto-optical Kerr effect (MOKE) in a setup equipped with an electromagnet capable of producing magnetic fields of up to 20 kOe in longitudinal geometry at room temperature. A 40 mW 409 nm laser was used with an angle of incidence of 14° with respect to the substrate plane. The initial beam was linearly polarized perpendicular to the plane of incidence (S-polarization) and modulated by a photoelastic modulator at f = 50 kHz. The second harmonic component, 2f, proportional to the angle of the Kerr-rotation, was extracted by a lock-in amplifier from the intensity of the reflected light.

Second harmonic generation was measured at room temperature using an 800 nm Ti-sapphire laser at a repetition rate of 82 MHz, the pulse average power and duration were 5 - 30 mW and 100 fs, respectively. The fundamental beam, whose polarization was varied with a half-wave plate, was focused onto the sample with the minimal cross-section of about 20 μm at the angle of incidence of 45°. SHG polar dependence was measured with the analyzer fixed at a polarization parallel (P) to the plane of incidence, and with the substrate edges fixed parallel to the analyzer position (=at 45 degrees to the in-plane axis of the polar domain). The input polarization was rotated continuously by $360^0$ degrees starting from polarization parallel (P) to the plane of incidence. A schematic of the geometry used for this experiment is presented on Figure S12.

The piezoelectric properties of the samples were characterized using a modified piezoelectric force microscopy (PFM) technique[21]. The samples were patterned with interdigitated electrodes using optical lithography, forming an alternating sequence of electrical ground and potential lines. The electrodes were Ti 20 nm / Au 80 nm with a width and spacing of 30 μm each. For the PFM-based measurements, the voltage was applied directly to the interdigitated electrodes and the lateral and vertical sample surface displacements were measured by monitoring the cantilever displacements using calibrated lock-in techniques. The sensitivities of the vertical and lateral signals were calibrated using a piezo stack which was in turn calibrated using an accurate laser Doppler vibrometer. For the lateral signal, the displacements measured in this work lie in the linear region, below the onset of tip-surface sliding, as we have verified. The vertical signal also lies in the linear region with indentation of the tip into the surface being negligible. For displacement profile measurements the excitation voltage had 1 V amplitude at 10 kHz. The measurement frequency was chosen away from any cantilever resonance. For the scanning mode, to increase the signal to noise ratio, the imaging was done at a cantilever resonance frequency, typically around 320 kHz. In this case, the PFM imaging is restricted to measurements of the vertical displacement.

All the crystal structures presented in this work were drawn using the program VESTA[22].

Superlattices were constructed in supercells based on the GdFeO$_3$ structure ($\sqrt{2}\mathbf{a_p}\times$ n $\mathbf{a_p}\times \sqrt{2}\mathbf{a_p}$, $\mathbf{a_p} \approx 3.9$ Å). Spin polarized calculations on Fe containing compounds were performed with G-type antiferromagnetic ordering consistent with the parent materials. The unit cell and atomic positions were optimized until the forces acting on each atom were less than 0.001 eV/Å. The fully relaxed structures were symmetrized using the FINDSYM programme[23] and the symmetric structures then further optimized to give a final structure. The resulting tight level of convergence was required to allow for an accurate calculation of the polarization for each structure.

Periodic density functional theory calculations were performed using the PBE functional[24] with the addition of an on-site Hubbard term for Fe with $U_{eff}$ = 4 eV[25]. A plane-wave cutoff energy of 550 eV was used, and a k-point grid of $6 \times 6 \times 4$ for the 1-1 structures (the number of k-points in the c direction was reduced according to the height of the supercell). The geometry optimization calculations were performed using VASP[26] and the projector augmented wave method[27].

Following geometry optimization, the static polarization was calculated in three ways. In the simplest method, the polarization was calculated using nominal static charges on each of the ions ($Ln^{3+}$, $Y^{3+}$, $Fe^{3+}$, $Ga^{3+}$, $O^{2-}$), and calculating their displacement away from a non-polar reference structure with Cmmm symmetry for odd and C2/m symmetry for even structures. Electronic effects were then included by using calculated Born effective charges in place of the static nominal charges. These were calculated for the final polar structure using density functional perturbation theory as implemented in VASP[28]. Finally, the polarization was also calculated for certain structures using the Berry phase method using the Quantum Espresso code[29] using the VASP optimised structure and equivalent DFT settings. These three methods are described in more detail in a recent review[30]. The relaxed-ion piezoelectric tensor was calculated using the implementation in VASP.

Classical force-field calculations were using the General Utility Lattice Programme (GULP)[31]. The force-field was constructed from a long range electrostatic part, which uses a shell model[32] to include a level of polarisability, and short-range Buckingham potentials[33] acting between pairs of ions, excluding cation-cation pairs. The parameters for the potential were chosen from previous literature[34, 35], and tested on bulk YFeO$_3$ and LaFeO$_3$. Experimental cell parameters for these orthoferrites were reproduced with a maximum error of 2 %. Partial occupancy of sites was treated using the mean-field approach in which the electrostatic and Buckingham potentials of that site are the average of the potentials for each species on that site, weighted by their occupancies.

## Results and discussion

**Symmetry considerations.** Tilting cannot produce polar symmetries in single A site perovskites[36]. However, in the presence of cation ordering on the A site along the $[010]_{Pnma}$ direction which is coupled to in-phase tilting of the octahedra along the same direction, a spontaneous electrical polarization arises.[13-16] This is an improper ferroelectric[37] as the primary order parameter is the zone boundary octahedral tilting. The polarization arises from non-cancellation of antiferrodistortive displacements in the $X5^+$ mode (of the parent cubic perovskite) at the interfaces between the blocks, producing a polarization density wave as seen in the lead-free ferroelectric $Bi_{0.72}La_{0.28}(Fe_{0.46}Ti_{0.27}Mg_{0.27})O_3$[38]. This observation can be further generalized in terms of the removal of inversion symmetry within the $X5^+$ mode present in tilted perovskites where there are perpendicular in phase and out of phase tilts, as highlighted in Figure 1 b. Indeed for single layer A site ordering Kishida et al.[17] have determined the symmetry of all possible tilt systems and cation ordering directions. Of these by far the most common in known perovskites is the $a^-b^+a^-$ tilt scheme producing the Pnma symmetry discussed above.

The generation of the sharp-interface A site alternation in single perovskite superlattices is challenging experimentally. We therefore used symmetry to generalize the crystal chemical rules defined previously[13-16] by extending the size of the distinct $ABO_3$ blocks and found that superlattices in which odd numbers of unit cells of both components are present retain the uncancelled displacements at the interfaces between the two components and are thus still polar in space group $P2_1ma$ (standard setting $Pmc2_1$) when the growth direction is parallel to the axis about which the in-phase tilting takes place. This generalization was achieved by describing the A site cation ordering perpendicular to either the $a^-$ or $b^+$ tilt in terms of modulated structures in superspace.[39-41] Cation ordering on the A site will increase the periodicity along one of the three pseudocubic perovskite subcell axes, according to the relative orientation of the ordering direction and the octahedral tilts. The resulting symmetries can be economically evaluated in a modulated structure description and are summarized on Figure 2. Selection of the Pnma **b** axis as the ordering direction produces a generalised supercell described by the $(0,k_y,0)$ reciprocal space wavevector which spans the $\Delta_1$ irreducible representation of Pnma and affords superspace groups described as Pnma$(0\beta0)$. Here the modulation vector $\beta$ is the selected value of $k_y$ and is the inverse of the multiplication of the $2\mathbf{a_p}$ repeat of the original cell along the ordering direction in real space. Orderings along the out-of-phase tilt axes are described similarly as $M_1(\alpha0\alpha)$ and $M_1(\alpha0\text{-}\alpha)$ affording $P2_1/m(\alpha0\alpha)$ and $P2_1/m(\alpha0\text{-}\alpha)$ superspace groups – here some symmetry elements of Pnma do not leave the wavevector invariant and hence the symmetry is lowered. The Brillouin zone and the definition of the k-vector types for the Pnma space group can be found on the Bilbao Crystallographic Server[42].

To describe the modulation in occupancy we define a fourth dimension $x_4 = t - \beta y$ (Figure 3a) in superspace, to define the atomic positions in terms of the fractional coordinates and occupancies of the Pnma subcell plus the effect of the modulation on these parameters on moving from one real space subcell to another (as the modulation has no component along $x$ or $z$, the position on $x_4$ is unchanged as we project cells into real space along these directions). If the structure is incommensurate the origin $t$ is arbitrary as all points on $x_4$ are projected. However when we consider commensurate structures, only certain points on $x_4$ are visited and the choice of $t$ is important for determining the real space structure. Independent of the exact nature of the modulation, the symmetry of the resulting structure can be determined from the origin $t$ and the modulation vector.[43, 44] Each choice of origin $t$ along $x_4$ corresponds to a specific real space A site occupancy and the full higher dimensional structure describes all possible combinations of A site occupancies within the periodicity defined by $\beta$.

The spatial variation of the A cation occupancy within the supercell can then be described by the periodicity of the ordering (defined by $\alpha$ and $\beta$ above) and the nature of the site occupancy variation. This is described by any periodic function of $x_4$: a crenel function (Figure 3c) of $x_4$ gives occupancies of zero or one at each A site,[45] and the width of the crenel function then defines the superlattice stoichiometry, with a width of 0.5 corresponding to a 50:50 ratio of the two different A site cations and a width of 0.6 corresponding to a 60:40 ratio. Figure 3 shows two possible origin choices for a 50:50 superlattice with $\beta = 1/5$ that is polar in $P2_1ma$ symmetry and a mixed superlattice that is non-polar in $P112_1/a$. Only these two types of origin choices are compatible with the crenel description here.

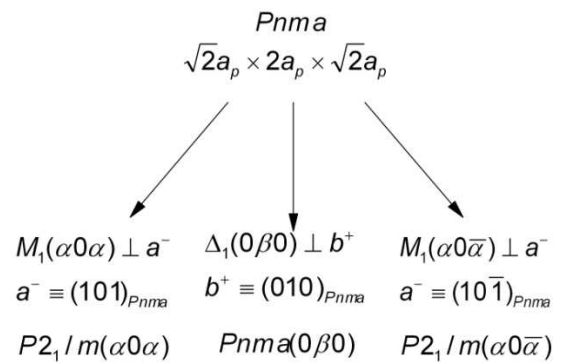

**Figure 2:** Layered A cation order perpendicular to the three Pnma axes can be treated as modulations at the $M(\alpha0\gamma)$ point, in the Brillouin zone for Pnma[42], for the $a^-$ axes (the general point corresponds to general planar orders perpendicular to the ac plane and the special cases $(\alpha0\alpha)$, $(\alpha0\text{-}\alpha)$ to the two planes of interest) and $\Delta(0\beta0)$ point for the $b^+$ axis, subscripts refer to particular irreducible representations which describe the modulation.

This can be generalized to describe any A site crenel-like superlattice such that for 50:50 lattices all $\beta = 1/(2N)$ structures with even numbers of unit cells are centrosymmetric whilst $\beta = 1/(2N+1)$ are polar. This result, i.e. odd-odd superlattices are polar, even-even ones are not holds for non 50:50 lattices with odd-odd or even-even combinations (see above for the 6:4 case). Odd-even cases are described by $\beta = 2N/(2N+1)$ and are all centrosymmetric. More complex stacking sequences are of course possible and their symmetries can be similarly deduced. A similar argument can be made for B site ordered crenels showing that for 50:50 lattices all $\beta = 1/(2N+1)$ structures are centrosymmetric whilst all $\beta = 1/(2N)$ are polar. Similar treatment of $P2_1/m(\alpha 0\alpha)$ and $P2_1/m(\alpha 0 -\alpha)$ superspace structures show that crenel based structures are non-polar $P2_1/m$ supercells.

The above approach can also be generalized to arbitrary modulation functions e.g. for simple trigonometric functions when the origin is chosen such that the centre of the layer represents the maximum in occupation- illustrated by the sine wave modulation functions in Figure 3 c- the same rules are obtained. Note that an arbitrary choice of origin for the sinusoidal modulation function will give a polar structure (P11a), though this is induced by cation order by analogy with multiple color lattices, rather than by removal of specific symmetry elements.

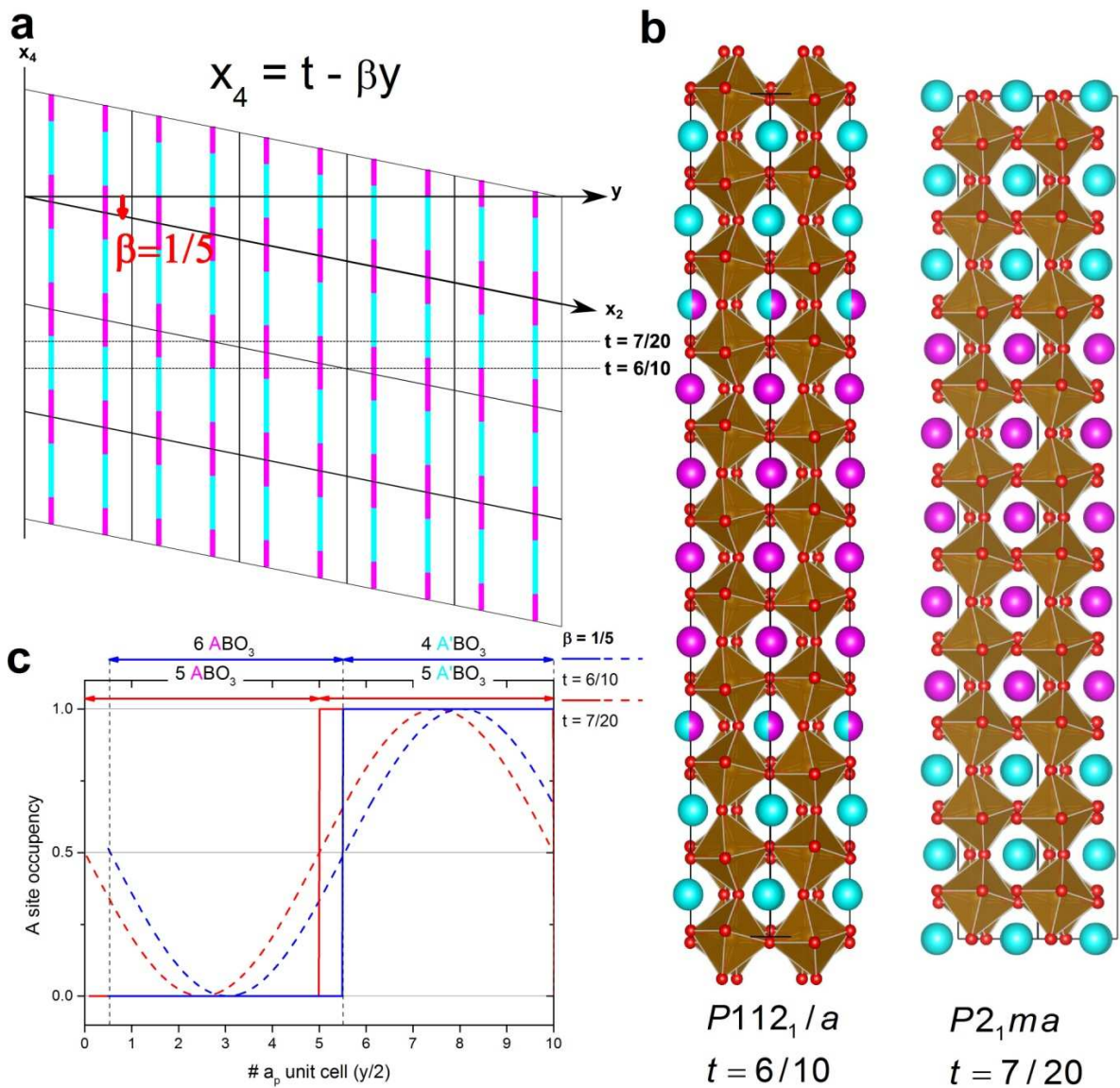

**Figure 3:** (a) Projection of the $x_4, x_2$ plane for $\beta = 1/5$ showing how the stacking sequence can be built up, here projections parallel to y correspond to structures in real space, where a cyan occupation domain is intersected a La is present on the A site and where a magenta one is intersected a Y is present. (b) and (c) Two example projections corresponding to origins of $t = 7/20$ (red functions in c) and 6/10 (blue functions in c) are shown. $x_2$ is the superspace coordinate related to y as defined in [39] such that the occupancy is periodic in $x_2$ for unit translations along y.

**Materials selection and computational analysis.** Materials selection began with the choice of $Fe^{3+}$ as the B site cation, as the strong $d^5$-$d^5$ superexchange interactions give high antiferromagnetic ordering temperatures and with A = $Ln^{3+}$ the resulting tilting of the octahedra due to tolerance factor considerations gives all $LnFeO_3$ perovskites the required Pnma symmetry. The resulting non-centric B-O-B superexchange pathways permit the formation of a weak ferromagnetic moment due to Dzialoshinsky-Moriya (DM) canting[46].

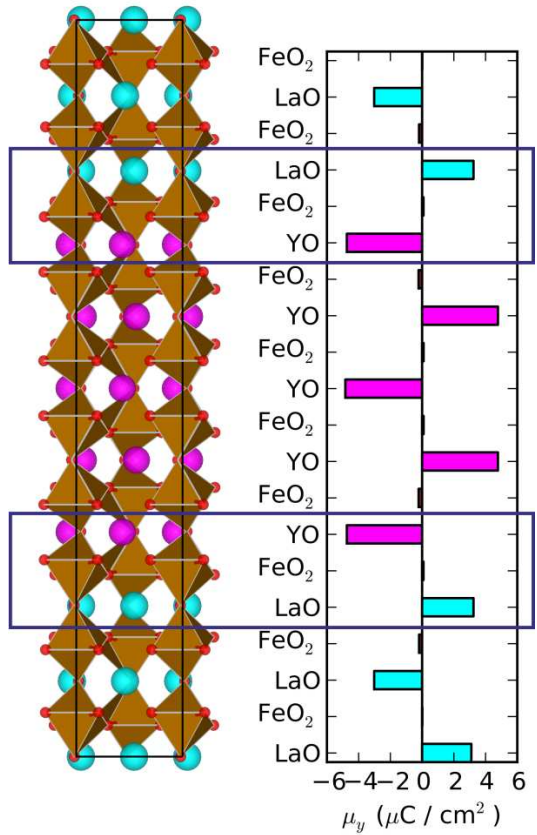

**Figure 4**: The calculated structure of the $(LaFeO_3)_5(YFeO_3)_5$ superlattice is shown alongside the contribution from each layer to the overall polarization. The box highlights the interface region where the local centre of symmetry is lost, leading to different contributions to the polarization from the AO layers above and below the interface and a finite overall polarization.

A consequence of the previous symmetry arguments is that for stacking of the individual $ABO_3$ and $A'BO_3$ components perpendicular to $b^+$, superlattices in which either component is present in even numbered blocks are not polar due to displacement cancellation (Figure S1), for stackings perpendicular to $a^-$ all superlattices are non-polar. The effect of the size of the Pnma building blocks on the total polarization was investigated computationally. These calculations showed that for odd numbered superlattices, extending the blocks expands the region between the $ABO_3$-$A'BO_3$ interfaces with almost non-polar bulk-like layers, such that the polarization is inversely proportional to the block thickness (Figures 4 and 5). These generalized rules allow some flexibility in achieving the required A site compositional modulation over these longer distances in heterostructures experimentally. Suitable $Ln^{3+}$ cations must be chosen to give polarity, and calculations show that for all odd numbered superlattices the polarization is maximized with the largest possible ionic size difference between the two A cations, leading to the selection A = La, A' = Y (Figure 5). This maximises the difference between the degree of displacement within AO and A'O layers. The final requirement is that the in-phase tilt axis, $b^+$, should be parallel to the stacking direction of the $LaFeO_3$ and $YFeO_3$ blocks. Calculations reveal that for $(LaFeO_3)_5(YFeO_3)$ superlattices only small energy differences separate the possible distinct tilt orientations, the $P2_1ma$ structure with the $b^+$ axis parallel to the stacking axis being less stable by only 0.06 eV/Formula Unit than the $P2_1/m$ structure with the $b^+$ tilt perpendicular to the stacking direction.

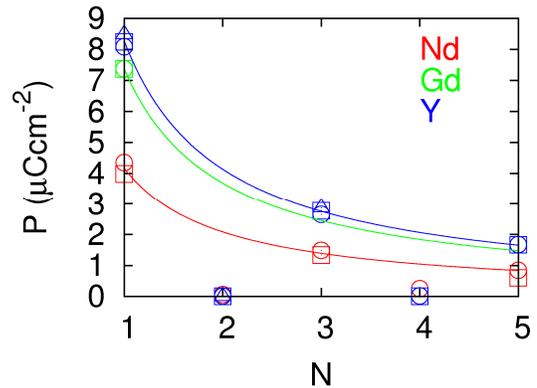

**Figure 5**: Polarization (P) versus the number of perovskite unit cells (N) in each block for $LaFeO_3$-$LnFeO_3$ superlattices (Ln = Nd, Gd, Y). Data is plotted for three methods of calculating polarization: static charges (squares), Born effective charges (circles) and Berry phase (triangles). The function P1/N, where P1 is the average calculated polarization for the 1-1 superlattices, is plotted as a line for each family.

**Growth.** Quasi-epitaxial growth at 600 °C of both selected orthoferrites on cubic substrates produces three orthorhombic domains corresponding to the different orientations of the $b^+$ tilt.[47-49] We observed layer-by-layer growth for $LaFeO_3$ on $SrTiO_3$ [100] (STO) above 600 °C (Supplementary Figure S5); below 600 °C the surface mobility is insufficient to induce crystallinity in the growing layers. We succeeded in growing $YFeO_3$/$LaFeO_3$ heterostructures on STO (Supplementary Figures S6 and S7). Higher quality heterostructures with fewer domains arise from growth on $DyScO_3$ [101] (DSO) (Figure S8). This substrate is a Pnma perovskite which can be treated to obtain a pristine $ScO_2$-terminated surface necessary for layer by layer growth[19] (Figure S3). The substrates used here have the $b^+$ tilt in-plane with two domains (Supplementary Figure S4) corresponding to the two orientations of this axis along the pseudocubic subcell directions. The good structural and dimensional match ($\varepsilon_s \approx 3.5\%$) allows YFeO$_3$ to grow layer-by-layer on DSO above 600 °C, producing a smooth two-dimensional surface after 5 unit cells are deposited, which permits subsequent growth of LaFeO$_3$ in layer-by-layer mode (Figure 6). Growth was carried out at the lowest temperature which afforded the RHEED oscillations characteristic of layer-by-layer growth to minimize A cation interdiffusion between the blocks.

**Structural characterization.** The periodicity of the superlattice is confirmed by the presence of a low-angle reflection in the specular x-ray reflectivity corresponding to a unit cell length of 38.8 Å showing that the structure is composed of 10 unit cells of the primitive perovskite (Figure 7 a). The observation of appropriately spaced satellite reflections around the fundamental LnFeO$_3$ perovskite out of plane reflections at higher angle (Figure 7 b), confirms that the structure is composed of 5 unit cells of LaFeO$_3$ and 5 unit cells of YFeO$_3$. It is important to note that only first order superstructure reflections from the $10a_p$ cell are seen, which is consistent with a non square-wave modulation of the occupancies of the A sites between successive blocks. Interdiffusion during growth produces a sinusoidal modulation of the A site occupancy away from the centre of the blocks which will not produce harmonic diffraction features.

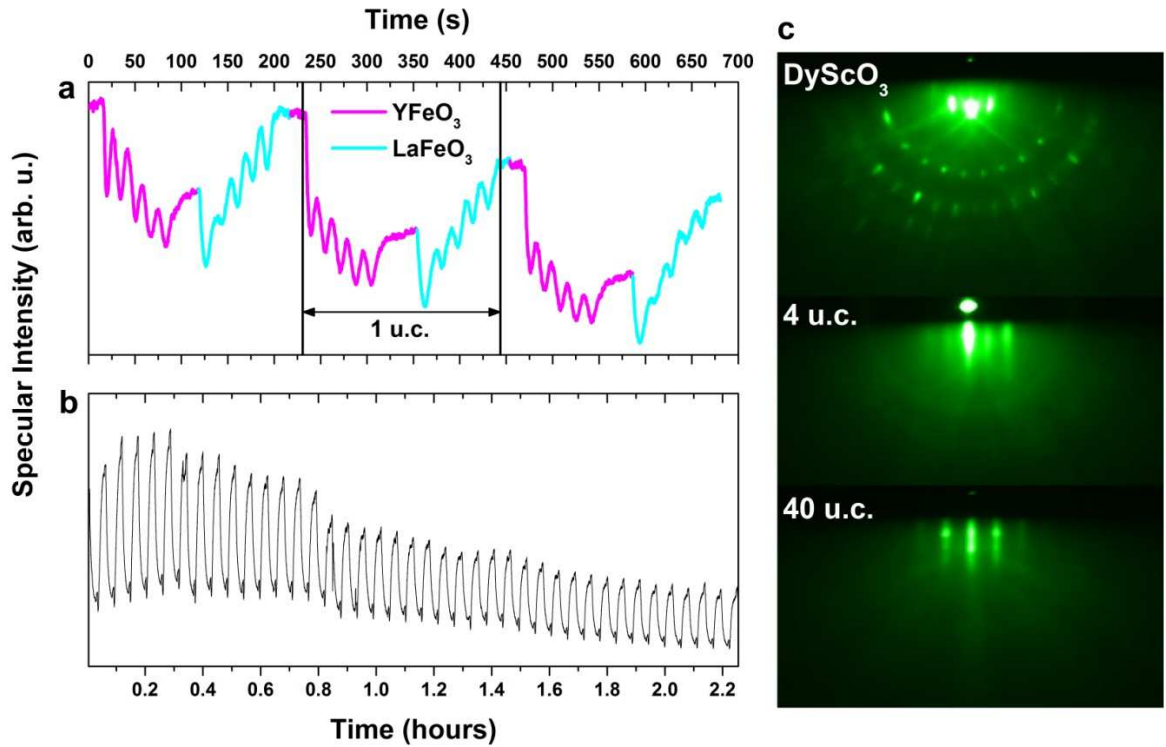

**Figure 6:** (a) Specular RHEED intensity as a function of deposition time for the first 3 unit cells showing 5 clear oscillations for both YFeO$_3$ and LaFeO$_3$ layers. (b) Specular intensity variation for the full heterostructure showing that the layer by layer growth is sustained for the 40 unit cells with generally decreasing intensity for the YFeO$_3$ layers and increasing intensity for the LaFeO$_3$ layers. (c) RHEED pattern of (top) ScO$_2$ terminated DSO substrate (middle) 4 unit cells and (bottom) 40 unit cells of the heterostructure along the [110] pseudocubic azimuth. The streaky pattern shows that the growth stays two-dimensional throughout the deposition.

This is confirmed by HAADF-STEM imaging which shows some interdiffusion across the interfaces which nonetheless remain clearly present in the images (Figure 8). The brightness in the projected columns is directly related to the average Z-number of the columns. The intensity profile, taken over the full width of the image, only displays maxima due to the A-cations, as those from the lighter B-cations (Z=26) do not rise above the base of the A-cation peaks. The alternation between higher intensity peaks corresponding to La-rich areas (Z=57) and lower intensity peaks from Y-rich areas (Z=39) follows a sine wave, i.e. the transitions between La- and Y-rich areas occurs gradually consistent with the superstructure intensity variation in the x-ray diffraction.

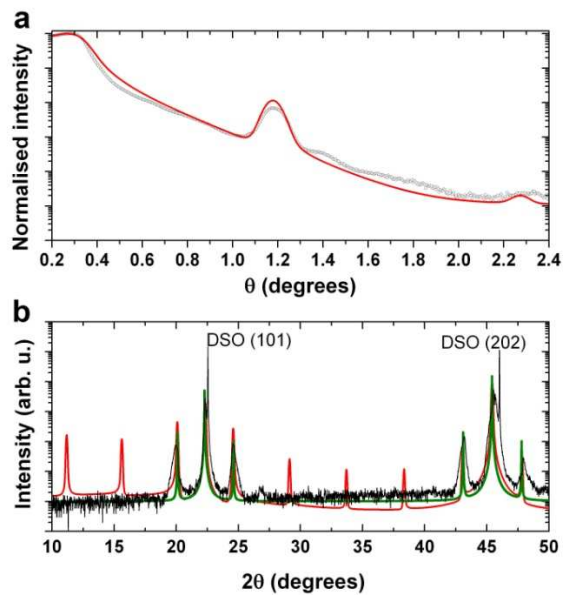

**Figure 7:** (a) x-ray reflectivity of the [(YFeO3)$_5$(LaFeO$_3$)$_5$]$_{40}$ heterostructure showing a Bragg reflection corresponding to an interatomic separation of 38.8 Å, the red line is a simulation of the heterostructure, the second feature around 2.3 degrees can be modeled using a surface roughness of 0.8 nm. (b) Out of plane θ-2θ XRD pattern of the heterostructure, the red line corresponds to a simulation of the structure using a crenel modulation (perfect interfaces) with the expected odd harmonic reflections whereas the green line corresponds to a simulation using a sinusoidal modulation of the A cation composition.

Classical force-field calculations were carried out in order to assess the effect of the experimentally observed interlayer mixing between A-sites on polarization. To test the reliability of the force-field the force-field optimized structure of the fully ordered (LaFeO$_3$)$_5$(YFeO$_3$)$_5$ superlattice with atomically sharp interfaces was obtained, and its polarization calculated using static charges. The structure and resulting polarization of 2.5 μC.cm$^{-2}$ were sufficiently close to those calculated with DFT (P = 1.7 μC.cm$^{-2}$) to consider the force-field as reliable. A (LaFeO$_3$)$_5$(YFeO$_3$)$_5$ superlattice was then constructed in which the A-site occupancy varied sinusoidally from a pure YO layer to a pure LaO in a period of five layers (Figure S2). Non-integer occupation of the A-sites was treated using a mean-field approach in which the cation potential is the occupancy weighted average of that for Y$^{3+}$ and La$^{3+}$. The polarization of the modulated structure was then calculated using static charges, with an overall polarization of 2.4 μC.cm$^{-2}$, only slightly less than that of the fully ordered superlattice. The symmetry arguments above show that the polarity does not depend on the nature (sine wave or crenel) of the modulation, only on its periodicity and composition.

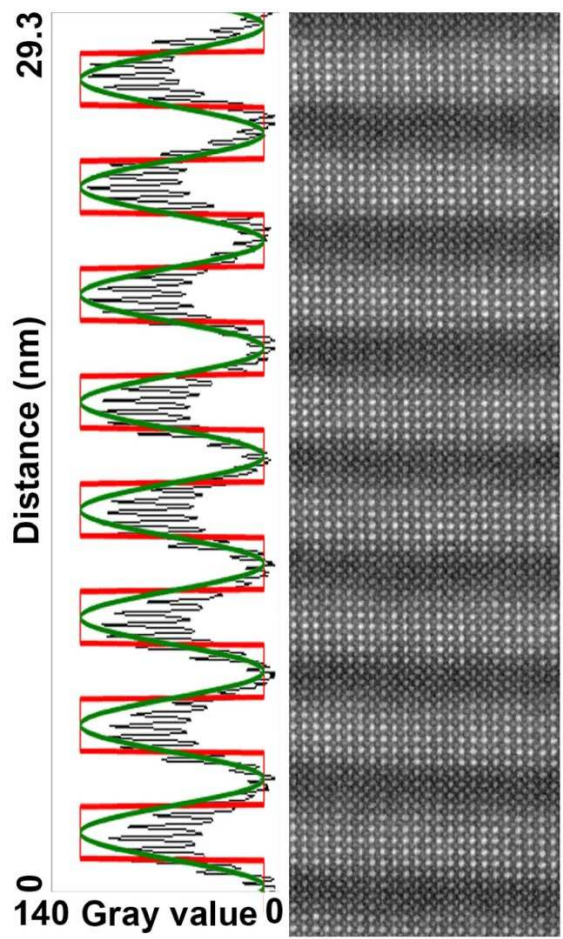

**Figure 8:** HAADF-STEM image (right) and intensity profile (left) over the same region of the [(YFeO$_3$)$_5$(LaFeO$_3$)$_5$]$_{40}$ heterostructure. The red and green functions superimposed on the intensity profile correspond to the crenel and sine-wave modulation models used to produce the XRD simulations. The gradual overall decrease in peak height from top to bottom is due to varying cross-section thickness.

Low temperature growth of orthorhombic LnFeO$_3$ films affords domains with the b$^+$ tilt both in- and out of plane[47-49]. The relative orientation of the in-phase tilt to the A site ordering is critical in determining the presence of polarity – when the in-phase tilt is aligned with the A site ordering, the symmetry is polar P2$_1$ma, whereas alignment of the out-of-phase tilt along this direction affords non-polar P2$_1$/m (Figure 3 d). In order to identify the relative orientation of the b$^+$ tilt the pole figure of the (111) reflection of the Pnma orthoferrite subcell was measured. This reflection of the perovskite subcell allows for the differentiation of the b$^+$ tilt orientation between polar P2$_1$ma and non-polar P2$_1$/m (Figure 9 d) structures. The comparison between the measured pole figures (Figure 9 a) and the expected pole figures for untwinned crystals (Figure 9 b and c) show that both tilt orientations with respect to the cation ordering direction are present (confirmed by measuring reciprocal space maps around the (111) reflections, Figure S9), and thus both non-polar P2$_1$/m and polar P2$_1$ma structures are formed in an approximate ratio of 5/1 based on diffracted intensities. This would be expected to produce both polar and non-polar domains in the film, and is consistent with the small calculated energy differences between these two structures, which both grow although the polar one is not matched with the substrate tilt pattern.

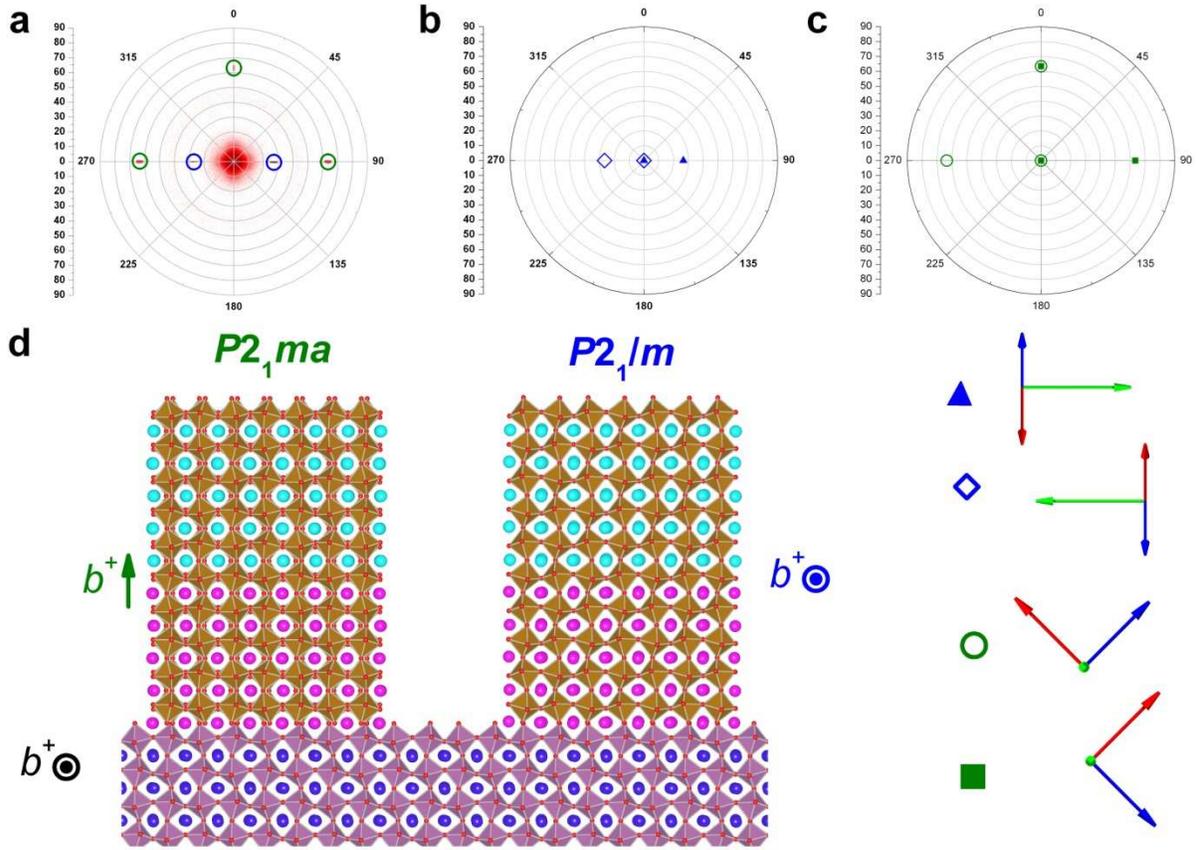

**Figure 9:** (a) Measured pole figure of the (111) reflection of the Pnma subcell showing two different domains corresponding to the non polar phase (peaks highlighted as blue circles) and the polar phase (peaks highlighted as green circles). (b) Simulated pole figure for the non-polar P21/m phase considering two variants corresponding to different orientations of the $2a_p$ $b^+$ axis. (c) Simulated pole figure for the polar P21ma phase considering two variants corresponding to different orientations of the $\sqrt{2}a_p$ $a^-$ axes. (d) Representation of the two phases obtained from the A cation ordering stacked along the $b^+$ tilt (left structure) and perpendicular to the $b^+$ tilt (right structure) together with the substrate. The key on the bottom right explains the relationship between the Pnma perovskite subcell orientations in the two domains seen for each of the two phases (the $2a_p$ direction is shown in green, the two $\sqrt{2}a_p$ directions in blue and red).

**Magnetic properties.** All the LnFeO$_3$ phases display weak ferromagnetism at all temperatures below $T_N$, except DyFeO$_3$ which is purely antiferromagnetic below 37K[50]. LaFeO$_3$ and YFeO$_3$ ($T_N$ 648K and 740K, respectively) both adopt the Γ4 magnetic structure with the weak ferromagnetic moment along the $2a_p$ $b^+$tiltPnma axis. The alternative Γ2 magnetic structure, where the canted moment lies along $b_{Pnma} = 2a_p$, has been observed in YFeO$_3$ above 70 kOe[51]. The sensitivity of the weak ferromagnetism in LaFeO$_3$ to the out-of-phase tilts via their control of the antisymmetric exchange and single-ion anisotropy[52] may produce deviations from the bulk magnetic structure in the heterostructure driven by interfacial and ferroelastic strains.

Absolute magnetization measurements on heterostructure grown on DyScO$_3$ substrates using standard magnetometry were not possible as the DyScO$_3$ paramagnetism dominates the film response.[53] Magnetization was measured on a superlattice grown on SrTiO$_3$ (Figure S10). The remnant in-plane magnetization is 0.011±0.001 $\mu_B$/f.u which is similar to the reported value for YFeO$_3$ films.[47] The in- and out-of-plane antiferromagnetic susceptibilities of the heterostructure were determined to be (11.3 ± 0.3) x 10$^{-7}$ and (5.4 ± 0.6) x 10$^{-7}$ $\mu_B$/Oe, respectively. These are of the same order found in YFeO$_3$ single crystals at room temperature,[50] demonstrating that the antiferromagnetic anisotropy strength in the heterostructure is comparable to the bulk properties of its components and thus that the heterostructure is magnetically ordered at room temperature. A qualitative characterization of the magnetic hysteresis loops has been performed using the magneto-optic measurements on a film grown on DSO substrates. Figure 10 a shows the MOKE loops of a (YFeO$_3$)$_5$/(LaFeO$_3$)$_5$ heterostructure with the magnetic field parallel to several high symmetry directions (Figure 10 d presents the relative orientation of substrate and films). We measured sharp hysteresis loops, similar to those observed on LaFeO$_3$ thin films of similar thickness grown under the same conditions (Fig. 10c). The analysis of the magnetic anisotropy of these films and the absence of weak ferromagnetism (WFM) when the magnetic field is applied parallel to the $[010]_{pc}$ axis confirms the $\Gamma 4$ magnetic configuration of the heterostructure films. The observed negative slope of the hysteresis loop for the heterostructure and LaFeO$_3$ is due to the contribution of domains with the $\mathbf{b_{Pnma}}$ axis out of plane (Equations S1 and S2) confirming the twinning (and the presence of polar and non-polar domains) described previously The degree of the negative slope is likely to be proportional to the number of structural domains grown with the $\mathbf{b_{Pnma}}$-axis out-of-plane.

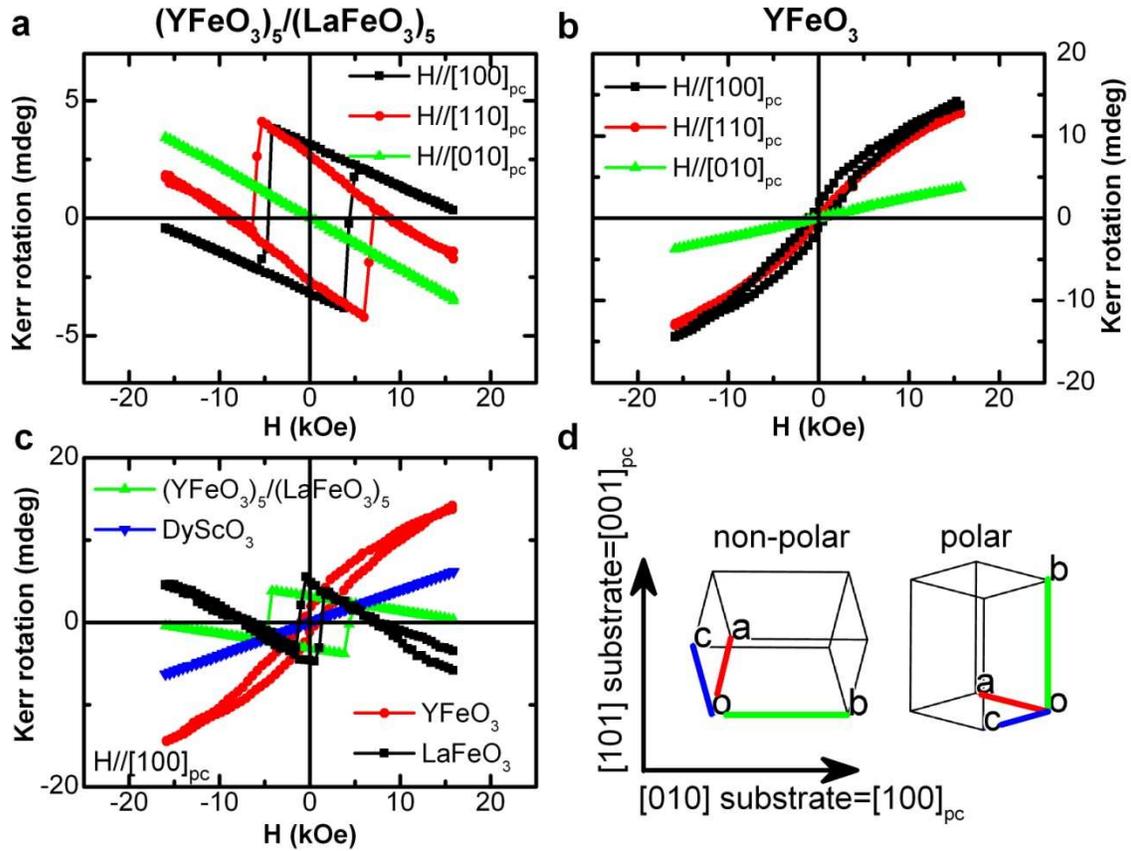

**Figure 10:** Magnetic hysteresis loops measured by longitudinal MOKE at room temperature on (a) the (YFeO$_3$)$_5$/(LaFeO$_3$)$_5$ heterostructure and (b) on thin YFeO$_3$ film both grown on DyScO$_3$ (101)$_{Pnma}$ substrates, with magnetic field applied in different directions with respect to DyScO$_3$ [010]$_{Pnma}$ or [100]$_{pc}$ in orthorhombic and pseudo-cubic notations, respectively. (c) Comparison of MOKE hysteresis loops measured on blank substrate DyScO$_3$ (101)$_{Pnma}$ and on thin films of (YFeO$_3$)$_5$/(LaFeO$_3$)$_5$, YFeO$_3$ and LaFeO$_3$ grown on DyScO$_3$ (101)$_{Pnma}$ substrates, respectively. The magnetic field is applied along the DyScO$_3$ [010]$_{Pnma}$ axis. (d) Three-dimensional drawings of orthorhombic unit cells corresponding to the polar and non-polar phases in (YFeO$_3$)$_5$/(LaFeO$_3$)$_5$ and their orientations with respect to orthorhombic (Pnma) or pseudo-cubic crystal axes of DyScO$_3$ (101)$_{Pnma}$ substrate.

**Non-linear optical properties.** To confirm the broken spatial inversion symmetry and to verify the polar point group of the 5:5 heterostructure we performed measurements of the optical second harmonic generation (SHG) polar plots on $[(YFeO_3)_5(LaFeO_3)_5]_{40}$ heterostructure and on a blank $DyScO_3$ substrate, respectively. SHG provides a necessary condition for polar dipolar order[54]. Figure 11 shows the dependence of the SHG $I^{2\omega}(\theta)$ intensity vs. polarizer rotational angle. Numerical analysis using equation S3 agrees with the point symmetry group 2mm, of the $P2_1ma$ space group of the polar phase. Measurements on a blank $DyScO_3$ substrate yielded an 8 times smaller absolute SHG response and agreed with the non-polar group mmm expected for Pnma, confirming the polar nature of the heterostructure, similar to the data reported for strained $EuTiO_3$.[8]

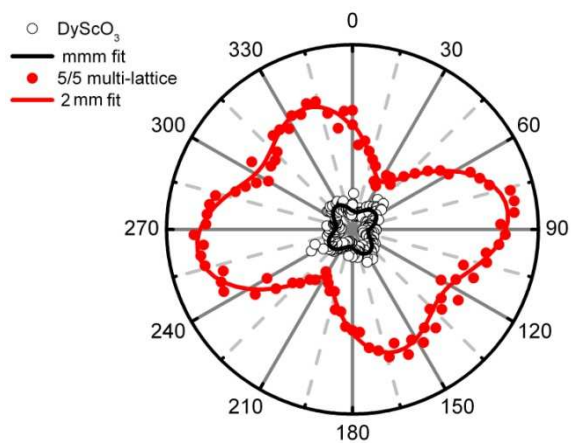

**Figure 11:** Optical second harmonic response vs. rotational angle of the polarizer. Filled (open) symbols and red (black) solid line represent data points and the corresponding fit for the film and blank $DyScO_3$ $(101)_{Pnma}$ substrate with point symmetry group 2mm and mmm, respectively.

**Piezoelectric force microscopy.** The functional nature of the structurally generated polarization was demonstrated by measuring the piezoresponse between interdigitated electrodes with the cantilever parallel to the electrodes (Supplementary Figure S13). The lateral and vertical displacement amplitudes are plotted in Figure 12 a and b with the negative sign used to indicate 180 ° phase difference between the applied voltage and measured cantilever displacement. Changing the voltage offset used alters both the displacement amplitudes and signs, as expected from the electric field dependence of the piezoelectric coefficient.[55] In order to verify the piezoelectric nature of the response and the electric polarization direction, finite element simulations were performed for the measured sample geometry using the relaxed-ion piezoelectric tensor with a calculated $d_{33}$ of 25.88 pC/N for the $(LaFeO_3)_1(YFeO_3)_1$ superlattice. The electrical polarization lies in the sample plane, perpendicular to the cation ordering direction as expected from the $P2_1ma$ symmetry produced orientation of the $b^+$ tilt. $d_{33}$ for the heterostructures was estimated within an order of magnitude as 10 pC/N. The agreement between this estimate and simulation confirms the piezoelectric origin of the sample surface displacements (Figure 12 a). Measurements with the cantilever perpendicular to the electrodes give a lateral signal within the noise level, indicating negligible polarization in this direction. We performed control experiments on separate $[(YFeO_3)_4(LaFeO_3)_4]_{50}$, $YFeO_3$ and $LaFeO_3$ films grown on $DyScO_3$ under identical conditions to those used to grow the heterostructures and did not measure any detectable piezoelectric effect (Figures 12a and S14). Scanning-mode PFM imaging of the surface of the $[(YFeO_3)_5(LaFeO_3)_5]_{40}$ heterostructure reveals both polar (bright area) and non-polar (darker areas) regions (Figure 12 c) with the polar region covering 25 % of the surface. The sample surface is very smooth with a root mean square roughness of 1 nm and no correlation was observed between the topography and piezoresponse signals.

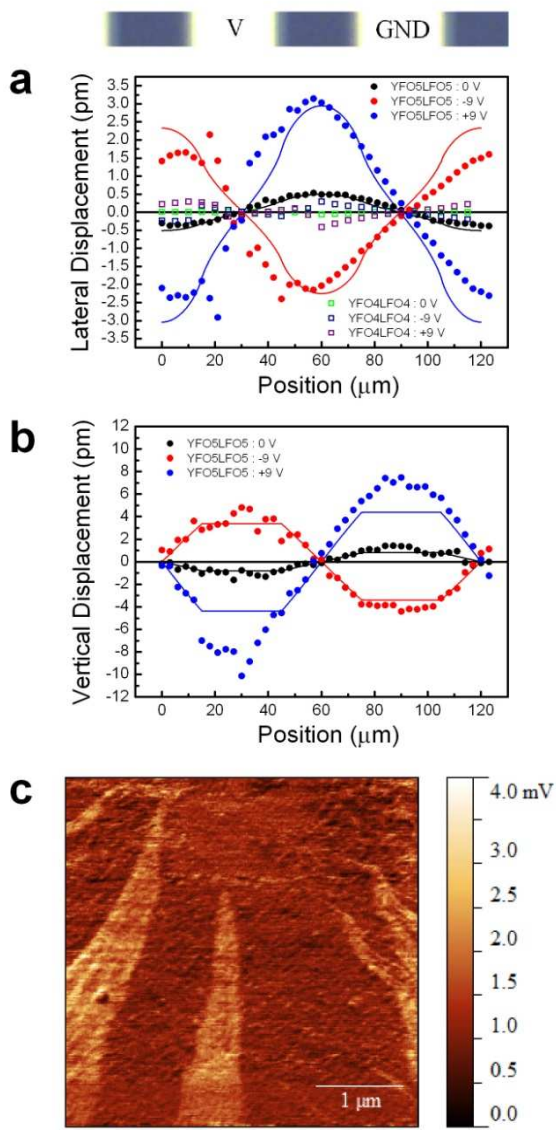

**Figure 12:** (a) Lateral and (b) vertical piezoresponse profile measurements (solid circles) and simulations (lines) for the $(LaFeO_3)_5(YFeO_3)_5$ and $(LaFeO_3)_4(YFeO_3)_4$ (square) thin films across a series of interdigitated electrodes shown in the top optical micrograph. The measurements and simulations are shown for three values of voltage offset, 0 V and ±9 V. (c) Scanning-mode PFM image of the $(LaFeO_3)_5(YFeO_3)_5$ film measured at a cantilever resonance using 1 V amplitude at 320 kHz. The vertical response amplitude is plotted, bright areas indicating polar domains whilst the dark areas are non-polar regions.

## Conclusions

$LaFeO_3$ and $YFeO_3$ are isosymmetric, non-polar and magnetically ordered at room temperature – spatial inversion symmetry is present, but time reversal symmetry is broken. Growth of alternating blocks of odd numbers of perovskite unit cells orders the A site $La^{3+}$ and $Y^{3+}$ cations. The cation order along the in-phase octahedral tilt removes the inversion centre in the component. The resulting uncompensated cation displacements from the centroids of their oxygen coordination polyhedra in the plane of the cation order produce a polarization perpendicular to the growth direction. The correct coupling of the translational (A site order) and point (octahedral in-phase rotation) symmetry is essential – if this is not achieved, non-polar structures result as shown by the presence of polar and non-polar regions in the PFM measurements. Rather than relying on a sharp compositional variation at the interface to break the symmetry, a more general approach is devised where coupled A cation ordering and specific mode distortions are the driving mechanism to create a polar system. This demonstrates the robustness of this approach to imperfection at the interfaces since a non-square-wave modulation of the composition will not affect the symmetry as long as the periodicity of the heterostructure remains correct. The A site occupancy modulation does not need to be square-wave to produce the polarization, it is the correspondence of the periodicity to an odd number of blocks of each component which is essential. The magnetic order of the component $AFeO_3$ blocks is retained in the heterostructure, with the structural polarity and functional piezoresponse imposed by the growth. The combination of magnetism and polarization at room temperature is thus achieved through the growth-controlled loss of symmetry. This mechanism does not require restrictive chemical criteria in the component ions ($d^0$ or $s^2$ configurations are not a pre-requisite, for example) and can be tuned via block size and chemical composition while still requiring only two distinct isosymmetric starting units, offering considerable potential diversity in property combination and control.

## Acknowledgements

This work is funded by the European Research Council (ERC Grant agreement 227987 RLUCIM). EDM, NES and NAI also acknowledge RFBR Grant #13-02-12450. Work done at WVU by PB and DL was supported by a Research Challenge Grant from the West Virginia Higher Education Policy Commission (HEPC.dsr.12.29).

## Notes and references

[a,] Department of Chemistry, University of Liverpool, Liverpool, L69 7ZD, UK
[b,] Stephenson Institute for Renewable Energy, Department of Physics, University of Liverpool, Liverpool, L69 7ZE, UK
[c,] National Physical Laboratory, Hampton Road, Teddington, Middlesex TW11 0LW, UK
[d,] Moscow State Technical University of Radioengineering, Electronics and Automation, Vernadsky Avenue 78, 119454 Moscow, Russia
[e,] EMAT, University of Antwerp, Groenenborgerlaan 171, B-2020, Antwerp, Belgium.
[f,] Department of Physics and Astronomy, West Virginia University, Morgantown, West Virginia, 26506-6315 USA.

Electronic Supplementary Information (ESI) available: Further details on computational methods, substrate preparation, thin film growth, structural characterization, magnetic measurement, nonlinear optical measurements, electrical characterization and piezoelectric force microscopy can be found in the Supplementary Information. See DOI: 10.1039/b000000x/

Supplementary Information for:

# Engineered spatial inversion symmetry breaking in an oxide heterostructure built from isosymmetric room-temperature magnetically ordered components

J. Alaria[a,b], P. Borisov[a,f], M. S. Dyer[a], T. D. Manning[a], S. Lepadatu[c], M. G. Cain[c], E. D. Mishina[d], N. E. Sherstyuk[d], N.A. Ilyin[d], J. Hadermann[e], D. Lederman[f], J. B. Claridge[a]*, and M. J. Rosseinsky[a]*,.

[a] Department of Chemistry, University of Liverpool, Liverpool, L69 7ZD, UK
[b] Stephenson Institute for Renewable Energy, Department of Physics, University of Liverpool, Liverpool, L69 7ZE, UK
[c] National Physical Laboratory, Hampton Road, Teddington, Middlesex TW11 0LW, UK
[d] Moscow State Technical University of Radioengineering, Electronics and Automation, Vernadsky Avenue 78, 119454 Moscow, Russia
[e] EMAT, University of Antwerp, Groenenborgerlaan 171, B-2020, Antwerp, Belgium.
[f] Department of Physics and Astronomy, West Virginia University, Morgantown, West Virginia, 26506 USA.

# 1 Calculations of polarisation for heterostructures of different thicknesses

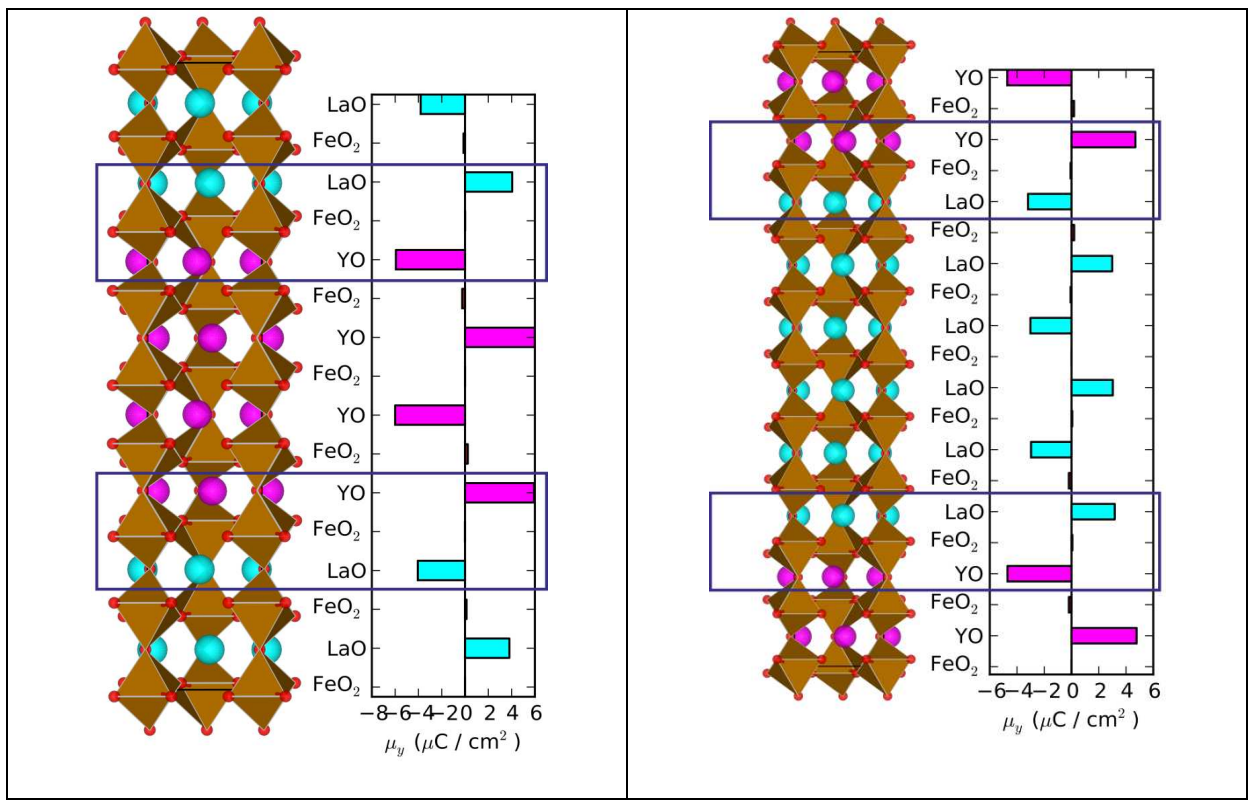

**Figure S1**. The calculated structure of the $(LaFeO_3)_4(YFeO_3)_4$ (left) and $(LaFeO_3)_6(YFeO_3)_4$ (right) heterostructures are shown alongside the contribution from each layer to the overall polarization. The purple box highlights the interface region where the displacements cancel each other, leading to zero net polarization.

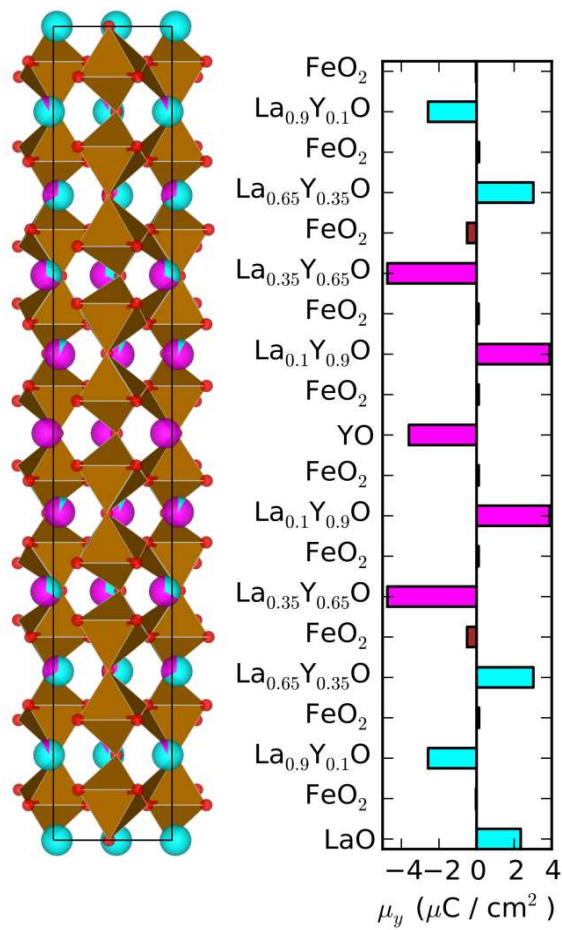

**Figure S2.** A structural diagram of the model used for the force-field calculation of $(LaFeO_3)_5(YFeO_3)_5$ with sinusoidal variation of the La and Y content along the stacking direction. The occupancy of the sites is represented by pie charts (La cyan, Y purple) on each atomic position. Plotted alongside is the layer-by-layer contribution to the polarisation in the polar direction calculated using static charges.

## 2 Substrate preparation and structure

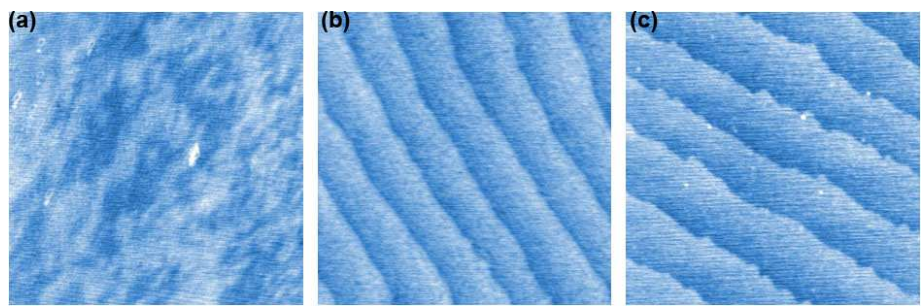

**Figure S3.** AFM of (a) as received, (b) annealed at $1050^{\circ}C$ for 2h in air and (c) etched for 1h in 12M NaOH/DI water solution DSO substrate showing the clear formation of an atomically flat surface

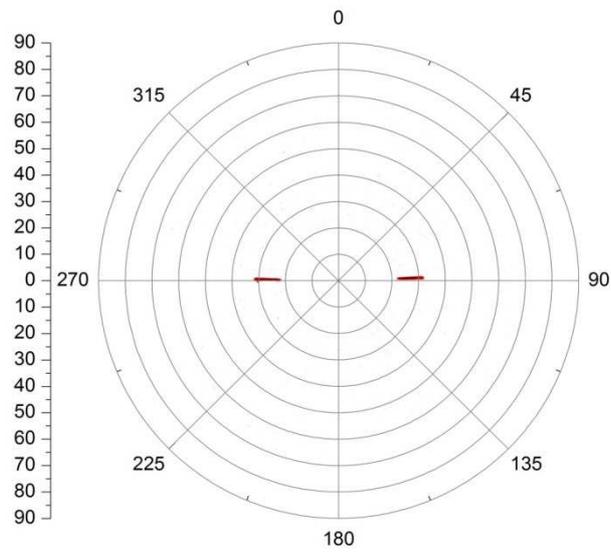

**Figure S4.** Pole figure on the (111)$_{Pnma}$ reflection of the DSO substrate showing a twofold symmetry, indicative of the two domains with the $b_{Pnma}$ in plane.

# 3 Thin film growth

We observed layer-by-layer growth for LaFeO$_3$ on SrTiO$_3$ [100] (STO) above 600 °C (Supplementary Figure S4); below 600 °C surface mobility is insufficient to induce crystallinity in the growing layers.

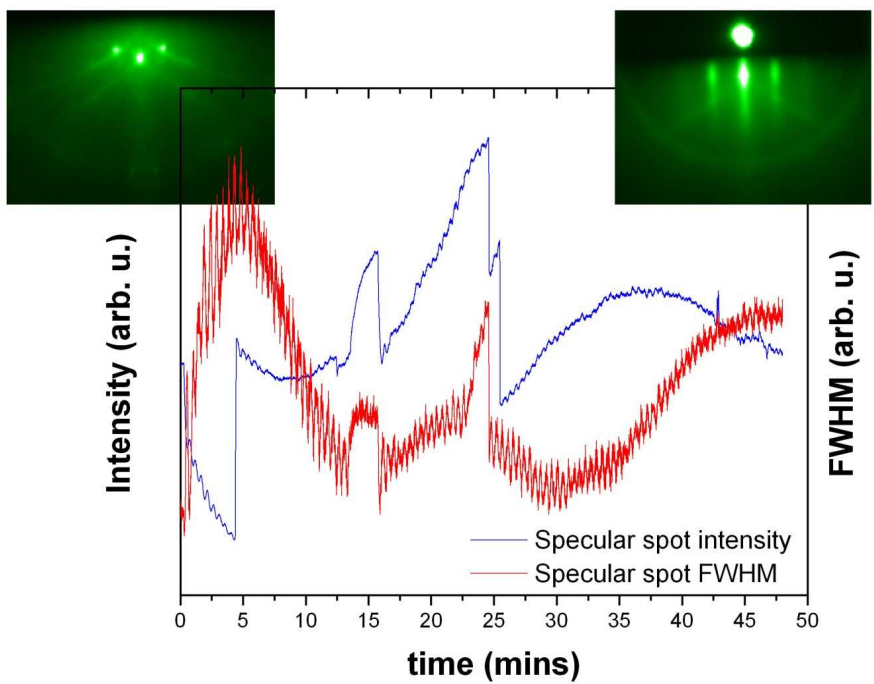

**Figure S5**. RHEED study of LaFeO$_3$ deposited in the same experimental conditions of the main text, showing sustained oscillation in both the specular intensity and FWHM of the specular spot

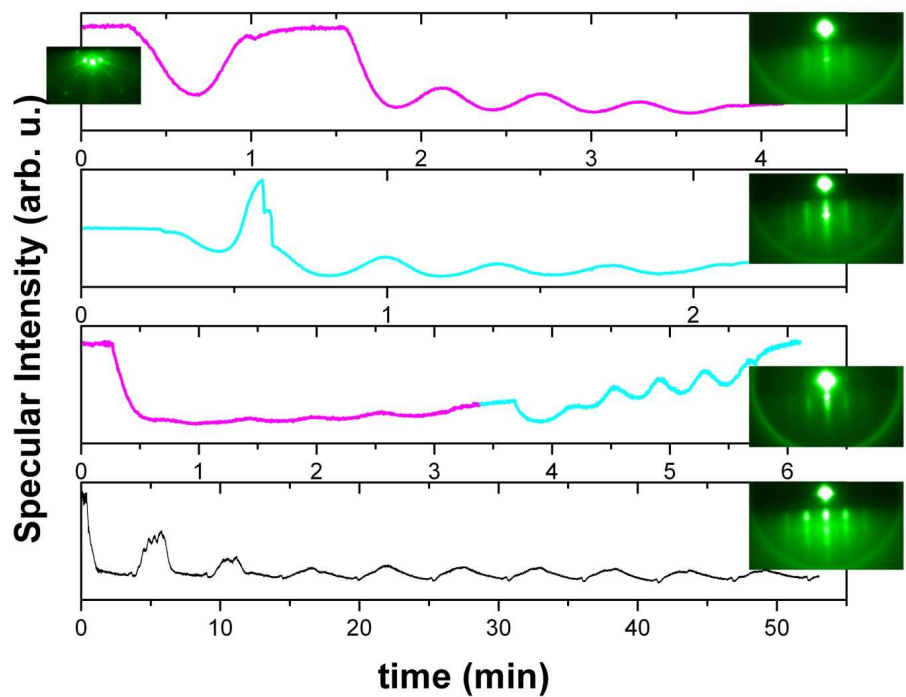

**Figure S6.** RHEED study of a $(LaFeO_3)_5(YFeO_3)_5$ heterostructure on STO showing the difficulty of obtaining clear RHEED oscillation for the YFO layer after a few monolayer. The RHEED pattern remains streaky indicating a 2 dimensional growth throughout the structure. The colour coding is identical to that in the main text (magenta for $YFeO_3$ and cyan for $LaFeO_3$ layer)

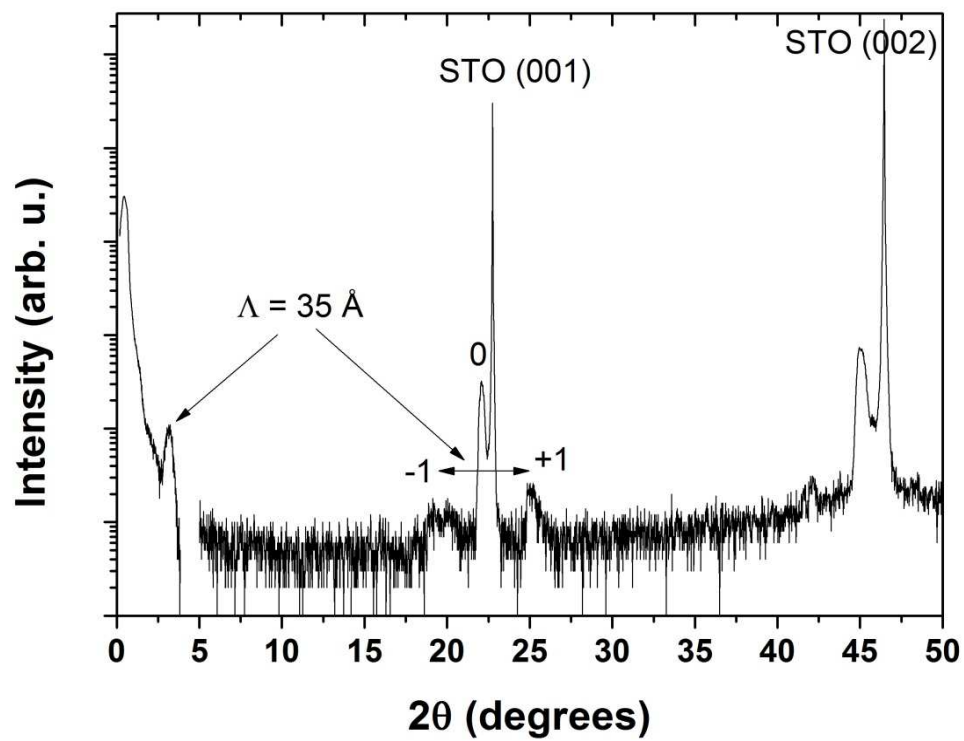

**Figure S7.** Out-of-plane θ-2θ X-ray Diffraction pattern for a $[(LaFeO_3)_5(YFeO_3)_5]_{40}$ grown on $SrTiO_3$ showing a Bragg peak at low angle consistent with a $10a_p$ ordering and weak satellite reflections around the main diffraction peak. The splitting of the peaks indicates the presence of different domains due to the difference of the out of plane lattice parameters.

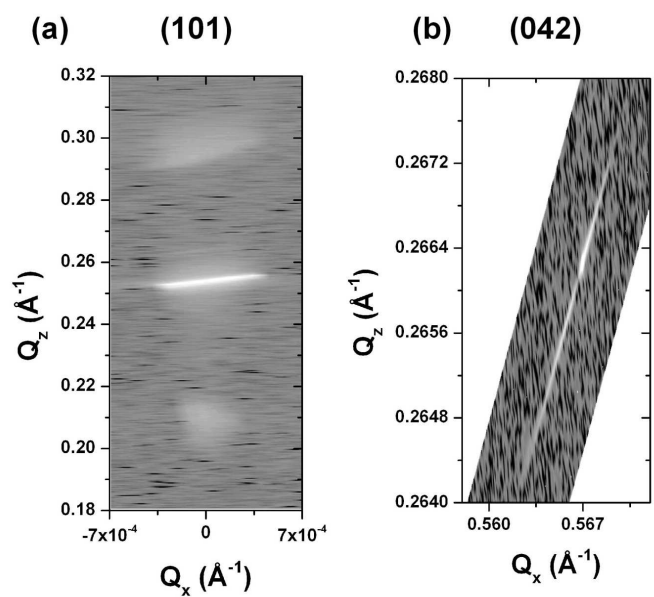

**Figure S8.** Reciprocal space maps of a [(YFeO$_3$)$_5$(LaFeO$_3$)$_5$]$_{40}$ heterostructure measured around (a) the (101)$_{Pmna}$ out-of-plane diffraction peak showing the good coherence between the heterostructure satellite and the substrate and (b) on the (042)$_{Pnma}$ subcell reflection showing both the substrate and the film peaks

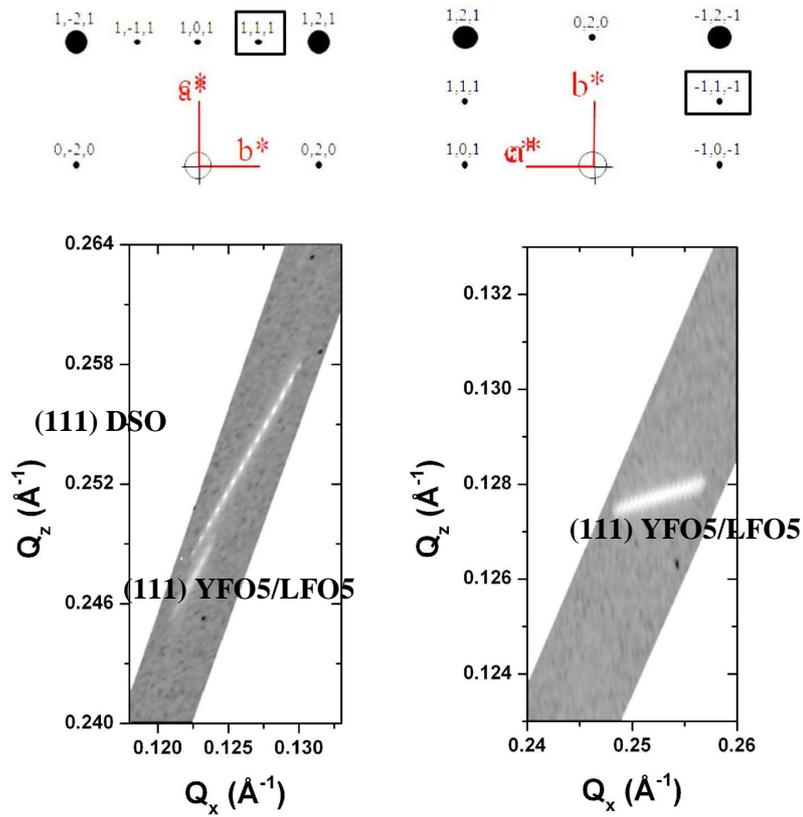

**Figure S9.** Reciprocal space maps of a [(YFeO$_3$)$_5$(LaFeO$_3$)$_5$]$_{40}$ heterostructure measured around the (111)$_{Pnma}$ subcell for (left) the non polar phase ($\chi=27°$) where both substrate and film peaks are observed and (right) the polar phase ($\chi=63°$) where only the film peak is observed. The simulated reciprocal space for each domain is shown on top with a black square showing which region is scanned.

# 4 Magnetic properties

## 4.1 SQUID Magnetometry

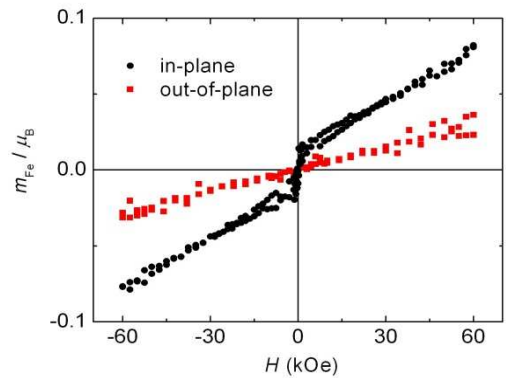

**Figure S10.** Magnetization hysteresis loops measured at T=300K on [YFO$_5$/LFO$_5$]$_{40}$ on STO (001) in magnetic field applied parallel (in-plane) and perpendicular (out-of-plane) to the film surface using a SQUID magnetometer.

Sample data have been corrected for the diamagnetic susceptibility of SrTiO$_3$ (100) substrates, measured at 300 K. Ferromagnetic contamination was removed by annealing the substrates used at 650 K in air for 2h as described by Yee *et al.*[1]

## 4.2 Magneto-optical Kerr Effect (MOKE)

Magneto-optical effects originate from magnetic perturbations of the dielectric permittivity tensor, where MOKE describes a change in the polarization of the reflected light, being linear proportional to the magnetization of the reflecting surface. MOKE has been extensively studied in orthoferrite materials[2]. Based on those studies a laser light source with the photon energy 3.0 eV was used, which activates the charge-transfer transition $t_{2u}^n(\pi) \rightarrow t_{2g}^*$ in $Fe^{3+}$ ions in octahedral crystal field of $O^{2-}$ ligands, thus producing one of the strongest maxima in the magneto-optic spectrum.

The longitudinal geometry (magnetic field applied parallel to both the film surface and to the plane of incidence) is primarily used to measure the magnetization components along the magnetic field, that is, in the film plane. However, the same geometry is also sensitive to the out-of-plane magnetization component if it does exist, as shown by numerical analysis of a similar MOKE setup.[3] The corresponding Kerr-rotation $\phi \propto a_1 m_x - a_2 m_z$, where $a_1$ and $a_2$ are parameters depending on material constants and angle of incidence. $m_x$ and $m_z$ denote the in-plane magnetization component along the applied field and the out-of-plane magnetization component, respectively. Because of the anisotropic orbital angular momentum quenching[2] the corresponding off-diagonal elements of the dielectric permittivity tensor, $\varepsilon_{ij}$, are proportional not only to the uncompensated magnetization, **m**, as usual, but also to the transverse antiferromagnetic vector **l**,[4]

$\varepsilon_{ac} = -\varepsilon_{ca} = i(\alpha_1 m_b + \alpha_2 l_a)$     (S1)

$\varepsilon_{bc} = -\varepsilon_{cb} = i(\alpha_1 m_a + \alpha_2 l_b)$     (S2)

where $\alpha_1, \alpha_2, \beta_1$ and $\beta_2$ are coefficients, and **m** and **l** are uncompensated magnetization and antiferromagnetic vector projections along the corresponding orthorhombic unit cell axis. Note that since the weak FM magnetization **m** is generated by AF spin canting, increasing the absolute value of **m** in the applied magnetic field would correspond to decreasing the AF vector projection **l** perpendicular to the field, and vice versa.

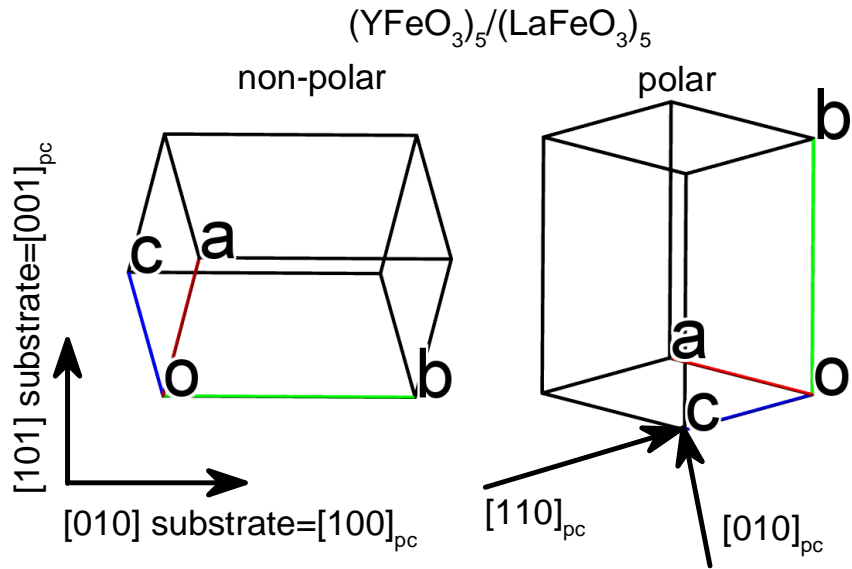

Figure S11: Three-dimensional drawings of orthorhombic unit cells corresponding to the polar and non-polar phases in $(YFeO_3)_5/(LaFeO_3)_5$ and their orientations with respect to orthorhombic (Pnma) or pseudo-cubic (pc) crystal axis of $(101)_{Pnma}$ substrate.

| (101)$_{Pnma}$ DyScO$_3$ | | (101)$_{Pnma}$ non-polar $(YFeO_3)_5/(LaFeO_3)_5$ | (010)$_{Pnma}$ polar $(YFeO_3)_5/(LaFeO_3)_5$ |
|---|---|---|---|
| pseudo-cubic (pc) | Pnma | Pnma | Pnma |
| [1 0 0] | [0 1 0] | [0 1 0] <br> $m_{WFM}$ in $\Gamma_4$ | [1 0 1] |
| [1 1 0] | [1 1 $\bar{1}$] | [1 1 $\bar{1}$] | [1 0 0] <br> $m_{WFM}$ in $\Gamma_2$ |
| [0 1 0] | [1 0 $\bar{1}$] | [1 0 $\bar{1}$] | [1 0 $\bar{1}$] |
| [0 0 1] | [1 0 1] | [1 0 1] | [0 1 0] <br> $m_{WFM}$ in $\Gamma_4$ |
| [0 1 1] | [1 0 0] | [1 0 0] <br> $m_{WFM}$ in $\Gamma_2$ | [1 1 $\bar{1}$] |

The expected MOKE responses depending on the magnetic structure ($\Gamma_2$ or $\Gamma_4$):

**Non-polar phase** (grown with the long axis b along [010] direction of the DSO (101) substrate):

If the magnetic field is applied along [010]=[100]$_{pc}$ of the DSO substrate, in orthorhombic and pseudo-cubic notation, respectively: the MOKE signal should contain weak FM hysteresis if the $\Gamma_4$ magnetic structure where the WFM moment lies along the b axis is adopted (compare Fig. 10a and 10b). When the magnetic field is along [110]$_{pc}$ of the DSO substrate, then the weak FM moment is oriented at an angle of 45 degrees to the magnetic field, that is, a hysteresis is possible, since the field projection along the easy weak FM axis [010]$_{Pnma}$ is still about 70% of the applied field (compare Fig. 10a and 10b, for magnetic data on single crystal of YFeO$_3$[5]. If the magnetic field is applied along [010]$_{pc}$ of the DSO substrate, then the weak FM moment is perpendicular to the field, and no hysteresis is expected. Since the transverse AF vector component would be zero, too, no contribution to the MOKE signal would be expected according to Eq. (S1) (compare Fig. 10a and 10b). Analog, under assumption of the $\Gamma_2$ magnetic structure one would expect a WFM hysteresis loop if the field is applied along [010]$_{pc}$ and no loop if the field is along [100]$_{pc}$. This is the opposite to the observed results (Fig. 10a). Based on the results shown in Fig. 10a, the magnetic configuration for the non-polar phase is $\Gamma_4$.

B: **Polar phase** (with the long axis b growing along [101] of the substrate):

If the magnetic configuration is $\Gamma_4$, then under all in-plane orientations of the magnetic field, the weak FM moment should be oriented perpendicular to the field (out-of-plane), therefore no FM hysteresis is expected. If the magnetic configuration of the polar phase is the alternative one, that is $\Gamma_2$, the weak FM magnetization should be in-plane and emerge in form of a FM hysteresis loop in case the field is applied along or at least at an angle of 45 degrees to the easy axis a (=along [110]$_{pc}$ of the substrate, see Fig. S11 and the table below. Since no FM hysteresis loop is observed in the latter case (Fig. 10a), when the field is applied along [010]$_{pc}$, the configuration $\Gamma_2$ is improbable. In the case of the $\Gamma_4$ configuration the hard axis (out-of-plane) magnetization as well as the corresponding transverse AF vector could still contribute to the total MOKE signal, however with the opposite sign if compared to the in-plane contribution. This explains the negative slope of the hysteresis loops shown for [(YFeO$_3$)$_5$/(LaFeO$_3$)$_5$]$_{40}$ and LaFeO$_3$ thin films grown on DSO presented in Fig. 10. The degree of the negative slope is likely to be proportional to the number of structural domains grown with the b-axis out-of-plane (=polar phase in the case of (YFeO$_3$)$_5$/(LaFeO$_3$)$_5$, see Fig. 10d). The experimental results shown in Fig. 10a are in agreement with the polar phase adopting magnetic configuration $\Gamma_4$. (YFeO$_3$)$_5$/(LaFeO$_3$)$_5$ heterostructures demonstrated (Fig. 10a) sharp hysteresis loops, similar to those observed on LaFeO$_3$ thin films (Fig. 10c). In comparison to those cases the YFeO$_3$ films showed similar magnetic anisotropy (Fig. 10b) but elongated hysteresis loops, which could be explained by the higher number of structural defects in those films due to the higher amount of the growth strain (-3.3%), that is, higher mismatch to DyScO$_3$ substrate, if compared to LaFeO$_3$ (mismatch -0.3%) and (YFeO$_3$)$_5$/(LaFeO$_3$)$_5$ (mismatch +0.1% and -0.2% for non-polar and polar phases, respectively).

Both systems, YFeO3 and LaFeO$_3$ films on DyScO$_3$ and (YFeO$_3$)$_5$/(LaFeO$_3$)$_5$ on SrTiO$_3$ showed similar coercive fields of about 1kOe, in agreement with the literature values[5] for in-plane magnetization in YFO films on STO substrates. In comparison to those, broader loops in the case of YFO-LFO films grown on DSO substrates with coercivity fields of about 5kOe could be related to the interfacial distortion in the polar phase.

## 5. Second Harmonic Generation
**Polar SHG dependences measured by polarizer rotation (angle θ) with fixed sample position:**

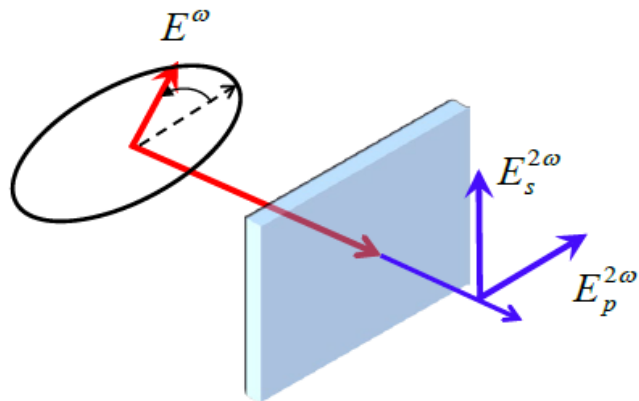

**Figure S12.** Schematic representing the geometry in which the rotational SHG dependence is measured with polarizer rotation (with the sample fixed) presented in the main text

The SHG intensity vs. θ curves in Fig. 12 were fitted using the following equation

$$I^{2\omega} = C_0 + C_2 \cos 2\theta + C_4 \cos 4\theta + S_2 \sin 2\theta + S_4 \cos 4\theta \tag{S3}$$

# 6. Electrical characterisation

The in-plane conductivity of the film was measured in a two probe configuration at room temperature using a electrometer (Keithley model 6430). The contacts were made using gold wires and indium solder. The conductivity was found to be $4.10^{-7}$ S/cm, the sample is therefore considered insulating at room temperature, a requirement to perform dielectric measurement.

## 7. PFM details

The piezoelectric properties of the samples were characterized using a modified piezoelectric force microscopy (PFM) technique. PFM is based on the standard contact mode atomic force microscopy (AFM) setup with the cantilever and tip being electrically conductive[6]. For piezoelectric samples a voltage applied between the tip and a bottom or surface electrode results in sample strains due to the inverse piezoelectric effect[7]. The sample strains cause vertical and lateral deflection of the cantilever which can be accurately measured using appropriate calibration methods. In order to separate the topography and piezoresponse signals and also to increase the signal to noise ratio a lock-in amplifier technique is used, with the voltage applied to the tip having a much larger frequency compared to the scanning frequency.

The profiles were measured with the cantilever parallel to the electrodes (Fig.S13) using lock-in amplifiers at 10 kHz with amplitude of 1 V. Various voltage offsets were used as indicated in Figure 5 d. The measurement frequency was chosen away from any cantilever resonance. The strain in the thin film underneath the electrodes is driven by the active regions between electrodes, resulting in a spatially alternating stretching/compression about their centre. At the same time, the thin film is displaced vertically, again alternating in sign between adjacent electrodes. By changing the voltage offset used, both the displacement amplitudes and signs can be changed. This is due to the electric field dependence of the piezoelectric coefficients[7]. As we have verified in Figure S13 (c) the displacement amplitudes change linearly with voltage offset, showing a non-hysteretic behaviour with coercive voltage of 2 V.

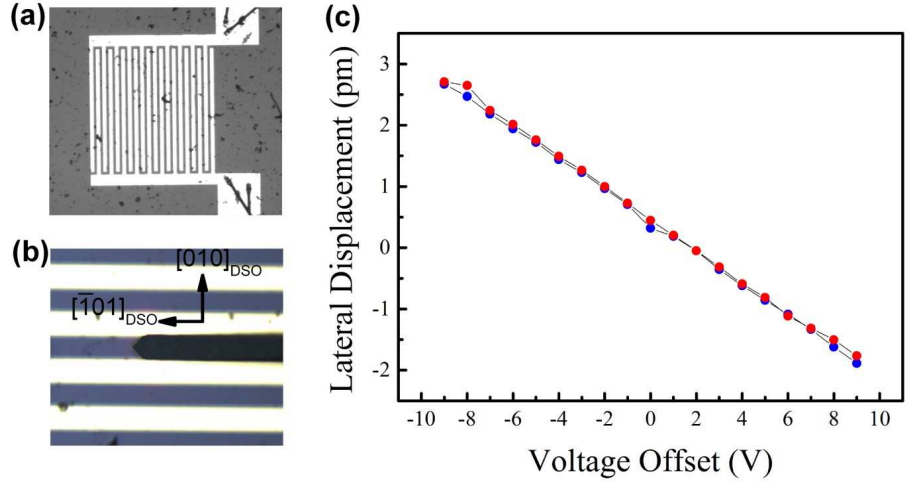

**Figure S13.** (a) Optical microscope images of the interdigital electrodes on top of the film (b) Respective orientation between the electrode and the *Pnma* axes of the DyScO$_3$ substrate (DSO). (c) Variation of the displacement amplitude as a function of the voltage offset.

The response of piezoelectric materials to electric fields and stresses is described by the system of strain-charge equations[7],

$$S = s^{Ec}T + d^T E$$
$$D = dT + \varepsilon^{Tc} E \qquad (S4)$$

where $T$ and $E$ are the stress and electric field vectors, respectively, $S$ and $D$ are the strain and electric displacement field vectors, respectively. $s^{Ec}$ denotes the compliance tensor for constant electric field, $d$ is the piezoelectric tensor and $\varepsilon^{Tc}$ is the permittivity tensor under constant stress. The piezoelectric tensor depends on the crystal structure and orientation, and thus also on the direction of the electric polarization. In order to verify the piezoelectric nature of the response shown in Figure 13 in the main text and the electric polarization direction, finite element simulations based on Equation 8 were performed for the measured sample geometry. The DSO substrate was modelled as a linear elastic material using literature values[8]. The back of the substrate was fully fixed in the simulations - in experiments the substrate was rigidly glued onto a metal plate. Moreover due to the interdigitated electrode structure long-range substrate deformations are negligible. The thin film heterostructure was modelled as a piezoelectric material, 160 nm thick, with a given polarization direction. Based on density functional theory calculation for the measured material we obtain the piezoelectric tensor

$$d/(pC/N) = \begin{bmatrix} 0 & 0 & 0 & -1.4 & 0 & 0 \\ 0 & 0 & 0 & 0 & 0.5 & 0 \\ -2.8 & -4.2 & 6.1 & 0 & 0 & 0 \end{bmatrix}$$

For the simulations, the polarization direction, and thus the piezoelectric tensor, is rotated to lie along a given direction, and the displacement profile was calculated for a potential of 1 V applied to a series of interdigitated electrodes, as in the measurements. The elastic properties of the thin film heterostructure depend on Young's modulus, $Y$, and Poisson's ratio, $\nu$, of the material. The best agreement between the simulations and the experimental data can be achieved by varying the ratio $\nu/Y$.[12] Within the order of magnitude, a good match between experiment and simulation is obtained for $\nu/Y = 10^{-12}$ $(m^2/N)$ and electric polarization perpendicular to the electrodes. This corresponds to the [-101] direction in the *Pnma* subcell coordinate system.

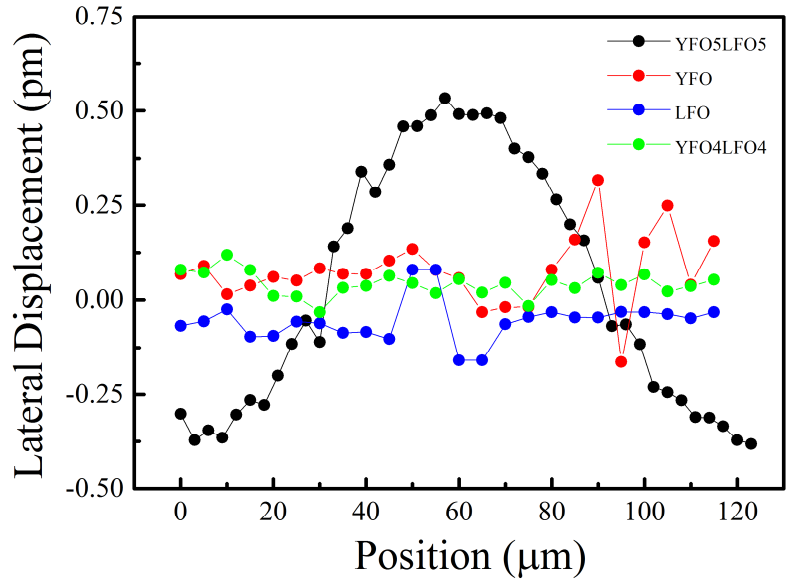

**Figure S14.** Lateral piezoresponse profile measurement on single layer of $YFeO_3$ (red), $LaFeO_3$ (blue) and $[(YFeO_3)_4(LaFeO_3)_4]_{50}$ grown on DSO (101) in the same condition as the heterostructure showing no piezoelectric effect within the limit of detection of the setup.